\documentclass[%
 reprint,
superscriptaddress,
 amsmath,amssymb,
aps,
prstab,
]{revtex4-1}
\usepackage{graphicx}
\usepackage{units}
\usepackage{tabularx,booktabs}
\usepackage{datetime}
\usepackage{color}
\usepackage[normalem]{ulem}
\usepackage{mathtools}

\renewcommand{\d}{{\rm d}}

\newcommand{\inj}{{\rm inj}}
\renewcommand{\th}{{\rm th}} 
\newcommand{\comb}{{\rm comb}}
\newcommand{\dB}{{\,\rm dB}}
\newcommand{\dBm}{{\,\rm dBm}}
\newcommand{\ns}{{\rm ns}}

\newcommand{\GHz}{{\rm GHz}}
\newcommand{\MHz}{{\rm MHz}}
\newcommand{\mA}{{\rm mA}}
\usepackage{tabularx,booktabs,soul}

\begin{document}

\title{Harmonic Frequency Locking and Tuning of Comb Frequency Spacing through Optical Injection}


\newcommand{\UCC}{Department of Physics, University College Cork, Ireland}
\newcommand{\Tyndall}{Tyndall National Institute, Cork, Ireland.}
\newcommand{\TUB}{Institut f\"ur Theoretische Physik, Technische Universt\"at Berlin, Germany.}

\author{Kevin Shortiss} \email{kevin.shortiss@tyndall.ie} \affiliation{\UCC} \affiliation{\Tyndall}
\author{Benjamin Lingnau}  \affiliation{\UCC} \affiliation{\Tyndall} \affiliation{\TUB}
\author{Fabien Dubois}  \affiliation{\UCC} \affiliation{\Tyndall} 
\author{Bryan Kelleher}  \affiliation{\UCC} \affiliation{\Tyndall} 
\author{Frank H. Peters}  \affiliation{\UCC} \affiliation{\Tyndall} 
\date{\today, \currenttime}

%
%
%


\begin{abstract}
We show, both experimentally and theoretically, that a slave laser injected with an optical frequency comb can undergo two distinct locking mechanisms, both of which decrease the output optical comb's frequency spacing. We report that, for certain detuning and relative injection strengths, slave laser relaxation oscillations can become undamped and lock to rational frequencies of the optical comb spacing, creating extra comb tones by nonlinear dynamics of the injected laser. We also study the frequency locking of the slave laser in between the injected comb lines, which add the slave laser's frequency to the comb. Our results demonstrate the effect of the $\alpha$ parameter, stability of the locked states, and indicate how the frequency of the relaxation oscillations affect both of these locking mechanisms. These optical locking mechanisms can be applied to regenerate or multiply optical combs. 
\end{abstract}



\maketitle 

\section{Introduction}

Optical frequency combs have been used in a myriad of applications since their realisation. Due to their high-precision, optical frequency combs (OFCs) have been applied in spectroscopy \cite{Bernhardt2010,Keilmann2004}, optical frequency metrology \cite{Udem2002}, optical ranging \cite{Trocha2018}, 
microwave frequency generators \cite{Fortier2011}, telecommunications \cite{Lundberg2018, Blumenthal2018}, 
among many other applications. While OFCs have seen increasing attention in the recent past, optical injection locking (OIL) has been an active area of research since the introduction of the maser\cite{Kurokawa1973}. Notable for its applications in increasing modulation bandwidths \cite{Simpson1997, Lau2008} and  reducing laser linewidth \cite{Wyatt1982, Chow1982}, optical injection also allows us to study the dynamics of non-linear oscillators, 
as the amplitude-phase coupling in a semiconductor laser's electric field can lead to complex dynamics.
Extensive literature detailing the behaviour observable in single frequency injection exists (see Ref. \cite{Wieczorek2005} and citations within), yet few studies have reported on the dynamics present when an OFC is injected into a single mode slave laser.

As OIL has been proposed as a method of demultiplexing OFCs in flexible optical networks \cite{Bordonalli2015, Zhou2015, Gutierrez2016, Cotter2019},
recent literature has been focused on how to optimise the side mode suppression \cite{ODuill2017, Shortiss2019} or stability \cite{Wu2012} obtainable when injecting optical combs into single mode lasers. 
Experimental and theoretical investigations into the dependence of side mode suppression on the detuning between the comb frequencies and the laser under injection have also been completed \cite{Wu2013, Gavrielides2014}. Ref. \cite{Gavrielides2014} acknowledges Hopf bifurcations can lead to relaxation oscillation (RO) excitations at high injection strengths, but focuses on the amplification seen by the unlocked lines, identifying several different regions with unique behaviour.

In this paper, we focus on two phenomena which in effect decrease optical the comb spacing due to the generation of additional comb lines. Firstly, we show the ROs in a slave laser are easily excited with optical comb injection and can be frequency locked to rational fractions $p/q$ of the optical comb spacing. The frequency locking of relaxation oscillations has previously been demonstrated in free-running Nd:YVO$_4$ lasers \cite{Otsuka1998}, where the relaxation oscillations associated with two separate laser modes were frequency locked to one another. In semiconductor lasers, it is known that single frequency optical injection can have a strong effect on the dampening and frequency of the ROs of a slave laser \cite{Mogensen1985, Petitbon1988, Hong1999}, and that ROs can be excited with single frequency injection \cite{Jagher1996}. To our knowledge however, frequency locking of the ROs in a semiconductor laser subject to optical injection has not previously been demonstrated. In Section \ref{Sec:Freq_Locking}, we first experimentally demonstrate RO locking to rational fractions $p/q$ of an injected optical comb spacing. A theoretical model is used to simulate the optical injection experiments, and good agreement is found between theory and experiment. Using our model, we further investigate the relationship between the RO locking width and the free-running RO frequency. We find that the locking width shrinks as the denominator $q$ grows in the rational fraction.

The second locking phenomena studied in detail in this paper are the additional frequency locked states observed as the slave laser is tuned in between the frequency lines  \cite{Lingnau2019, Tistomo2011}. These frequency locked states (also known as Arnol'd tongues \cite{Arnold1965}) appear as the slave laser is detuned by a rational fraction of the injected optical comb spacing. Using our theoretical model, we create a numerical map of the parameter space spanned by optical injection strength and detuning, illustrating the effective output optical comb of each frequency locked state. In Section \ref{sec:Stability_and_Alpha}, the stability of the frequency locked RO states and the resonance states within the Arnol'd tongues are investigated, with both types found to be very stable. The effect of the RO frequency and the magnitude of the amplitude-phase coupling $\alpha$ on the size of Arnol'd tongues in parameter space is also studied. We show that the nonlinearity induced by a large $\alpha$ enables locking of the ROs to higher order resonances of the injected comb, when the laser RO frequency is close to this harmonic comb spacing. This effect is largely suppressed for smaller values of $\alpha$, with only the second harmonic resonance remaining significant.

\section{Experimental Setup and Theoretical Model}
\label{Sec:Setup_and_Model}

\begin{figure*}
\centering
 \includegraphics[width=1.5\columnwidth]{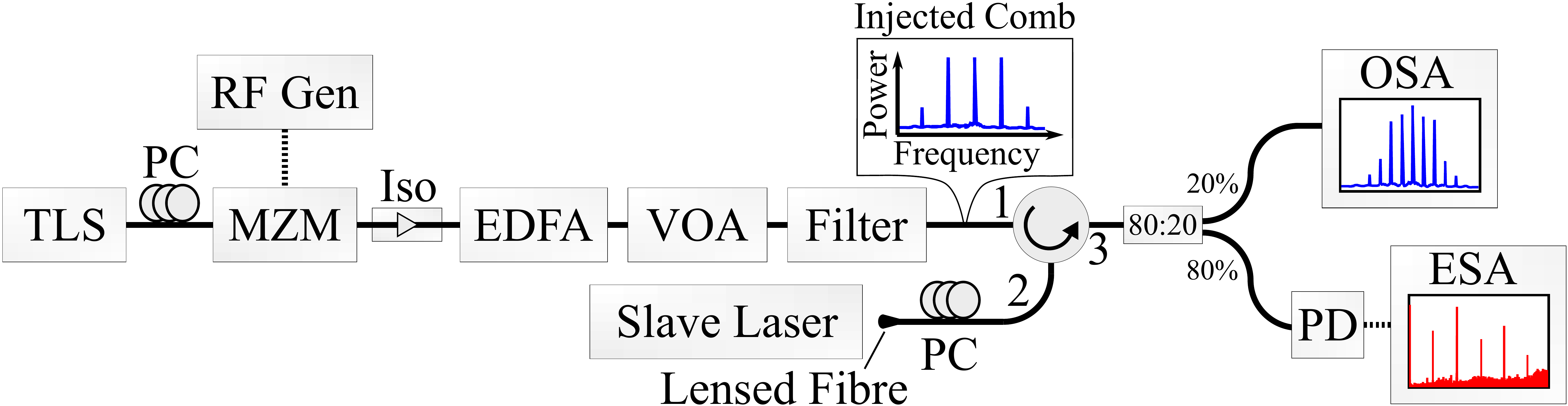}
 \caption{ 
Experimental setup for performing optical comb injection. TLS: Tunable laser source, MZM: Mach zehnder modulator, PC: Polarisation controller, EDFA: Erbium doped fibre amplifier, VOA: Variable optical attenuator, OSA: Optical spectrum analyser, PD: Photodiode, ESA: Electrical spectrum analyser.}
 \label{fig:Exp_Setup}
\end{figure*}

To study the frequency locking of the slave laser's relaxation oscillations (ROs) and lasing frequency under the injection of an optical frequency comb (OFC), a standard injection experimental setup was used, as shown in Fig.~ \ref{fig:Exp_Setup}. A commercial tunable laser source (TLS) was intensity modulated using a LiNbO$_3$ Mach-Zehnder modulator (MZM) to create a 3 line OFC. The comb's optical power was amplified and controlled using an erbium doped fibre amplifier (EDFA) and variable optical attenuator (VOA) in series. An optical filter removed a significant portion of the amplified spontaneous emission introduced by the EDFA. The slave laser used was an unpackaged single mode AlGaInAs quantum well device emitting at $1550~nm$, as described in Ref. \cite{Shortiss2018}. A lensed fibre was used to optically couple to the device, which is mounted on top of a temperature controlled brass chuck. A circulator was used to optically inject the slave laser, and the light output was monitored on an OSA and ESA simultaneously. 

In order to simulate the optical comb injection experiments described above, we formulate a rate equation equation model \cite{Ohtsubo2013}, modelling the dynamics of the complex electric field inside the cavity, $E(t)$, and the normalised active medium carrier density $N(t)$ (in units of inverse time).
(in units of inverse time).
\begin{align}
\frac{\d}{\d t} E = (1+ i \alpha) &\left(\frac{N-N_0}{2} - \kappa\right)  E + \frac{\partial E}{\partial t }  \Big|_{\inj} \nonumber \\
&+ \sqrt{\beta}~(\xi'_{sp}(t) + i \xi''_{sp}(t)) , \\ 
\frac{\d}{\d t}  N &= \frac{J - N}{T} - (N-N_0) |E|^2.
\end{align}
Here, $J$ is the normalised pump current, $N_0$ is the transparency carrier density, $T$ is the carrier lifetime, and $\alpha$ is the amplitude-phase coupling parameter. The optical cavity loss rate is given by $2\kappa$. We determine the parameter values from fits to measured RF spectra of the relative intensity noise of the free-running laser (as shown in Fig.~\ref{Fig_jsweep_with_RO_fits}). 

The optical comb injected into the laser cavity is modelled by an additional driving term:
\begin{align}\label{eq:injection}
 \frac{\partial E}{\partial t} \Big|_{\inj} &= K E_0 \Big( 1 + 2 m \cos(\phi_{\rm comb} ) \Big) e^{i\phi_\inj(t)} -  i 2\pi \nu_\inj E\,,
\end{align}
where $m$ is the relative strength of the injected side-modes at frequencies $\nu_\inj\pm\Delta$. We neglect comb modes further away from the center mode since, experimentally, their power is well below $-30\dB$ of the three strongest comb modes. The injection strength $K$ is the amplitude ratio of the injected field and the free-running laser emission, with $E_0$ the free-running intracavity field. The second term in Eq.~\eqref{eq:injection} transforms the electric field into the rotating frame of the central comb mode, at a frequency detuning of $\nu_\inj$ with respect to the free-running laser frequency \cite{Wieczorek2005}. 

\begin{table}
\centering
  \caption{Simulation parameters unless noted otherwise.}
 \begin{tabular}{cl l}
   \toprule[1pt]
  Symbol & Value & Meaning \\\midrule[1pt]
  $\kappa$ & $60~\ns^{-1}$ & Cavity field loss rate\\
  $\alpha$ & $3$ 		& Linewidth enhancement factor  \\
  $T$ 	   & $0.22~\ns$ &   Carrier lifetime\\
  $N_0$    & $100~\ns^{-1}$ & Normalised transparency carrier density\\
  $\Delta$ & $10~\GHz$ & Optical comb spacing\\
  $\beta$ & $10^{-5}~\ns^{-1}$ & Spontaneous emission strength \\
  $\Delta\nu_{\rm comb}$ & $1~\MHz$ & Optical comb RF linewidth\\
  $\Delta\nu_\inj$ & $1~\MHz$ & Master laser linewidth  \\
  \bottomrule[1pt]
 \end{tabular}
 \label{tab:parameters}
\end{table}

We model spontaneous emission noise by a stochastic forcing $\sqrt{\beta}(\xi'_{sp}(t) + i \xi''_{sp}(t))$, with the spontaneous emission strength $\beta$. Similarly, we implement a timing noise in the optical comb leading to an RF comb linewidth $\Delta\nu_{\rm comb }$, and a phase noise of the master laser, leading to an optical linewidth $\Delta\nu_\inj$:
\begin{align}
 \frac{\partial \phi_{\rm comb}}{\partial t} &= 2\pi\Delta + \sqrt{2\pi\Delta\nu_{\rm comb}}~ \xi_{\comb}(t) \\
 \frac{\partial \phi_\inj}{\partial t} &= \sqrt{2\pi\Delta\nu_\inj}~ \xi_{\inj}(t)
\end{align}
Here, $\xi_j(t)$ are complex Gaussian white noise source terms, fulfilling the correlation properties  $\langle \xi_j(t) \xi_k(t') \rangle = \delta(t-t') \delta_{j,k}$. The parameters used in the simulations are contained in Table \ref{tab:parameters} unless stated otherwise.
%

\section{Relaxation Oscillation Excitation and Frequency Locking}
\label{Sec:Freq_Locking}

\begin{figure*}
\centering
\includegraphics[width=1.8\columnwidth]{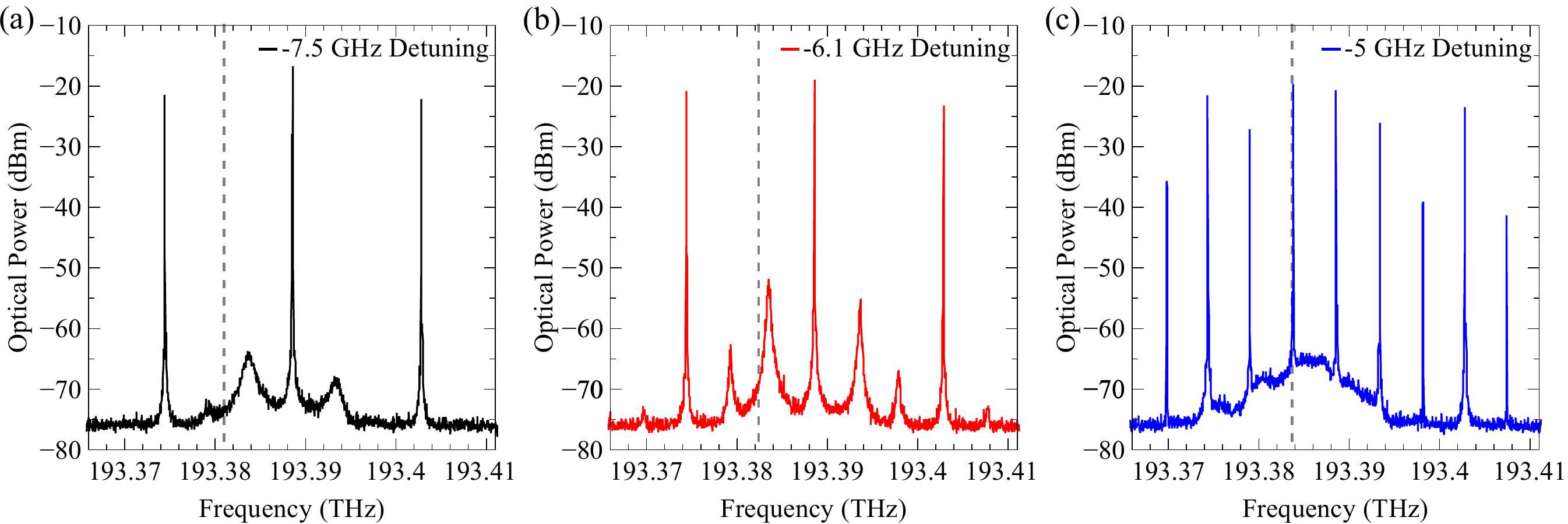}
 \caption{Optical spectra from a slave laser with $-4.2~\dBm$ free-running power, under injection from a $14~\GHz$  optical comb with optical power $+4.1~\dBm$.
In each case, vertical dashed lines indicate the free-running frequency of the slave laser. \textbf{(a)} Stable locking to the centre comb line, at $-7.5~\GHz$ detuning.  \textbf{(b)} At a detuning of $-6.1~\GHz$, relaxation oscillations becoming excited. \textbf{(c)} The excited relaxation oscillations lock to $\tfrac13$ the optical comb spacing, creating a new optical comb with spacing $4.66~\GHz$.}
 \label{fig:RO_Excitation_Plots}
\end{figure*} 

In the following section we present experimental and theoretical results which demonstrate two types of frequency locking. Both types of locking are distinct from the standard Adler-type locking found when an oscillator is driven with a single driving frequency \cite{Adler1946, Pikovsky2001}. 

Firstly, to demonstrate how the relaxation oscillations (ROs) of a slave laser become undamped under optical comb injection, in Fig.~\ref{fig:RO_Excitation_Plots} we present the optical spectra of a slave laser under injection of a three-line $14~\GHz$ comb
, at three different detunings. At a detuning of $-7.5~\GHz$ (Fig.~\ref{fig:RO_Excitation_Plots} (a)) the slave laser is stably locked to the centre line of the comb. The ROs in the slave laser are visible in the spectra, at $\pm 5~\GHz$ from the centre comb line and $-40~\dB$ below. As the detuning is decreased further in Fig.~\ref{fig:RO_Excitation_Plots} (b), the ROs begin to become undamped, rising in optical power. 
In Fig.~ \ref{fig:RO_Excitation_Plots}(c) the ROs lock to exactly $\tfrac13$ of the optical comb frequency spacing, generating further tones in the comb through nonlinear processes in the slave laser. In effect this decreases the optical comb spacing from $14~\GHz$ to $4.66~\GHz$. 

\begin{figure}
\centering
 \includegraphics[width=0.99\columnwidth]{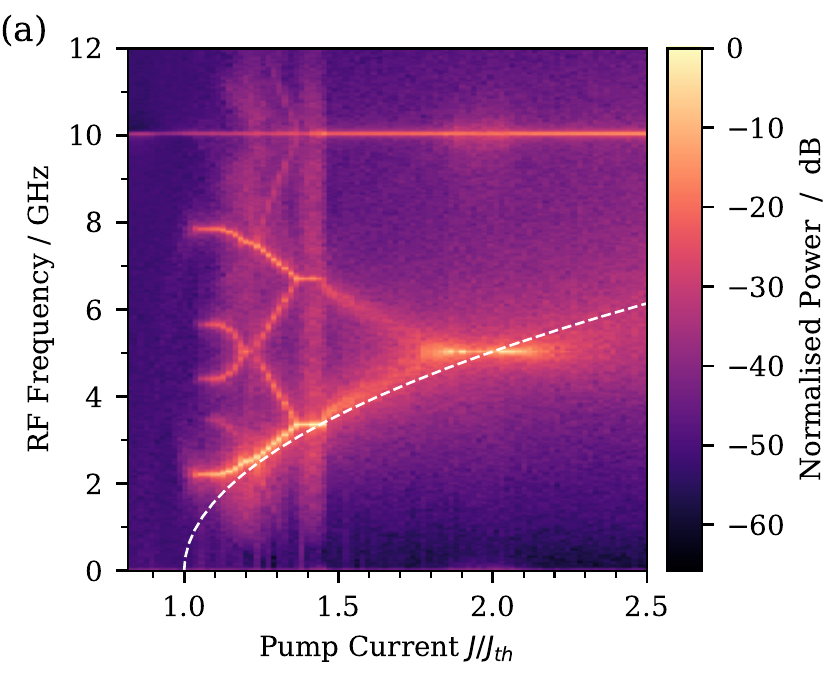}
 \includegraphics[width=0.99\columnwidth]{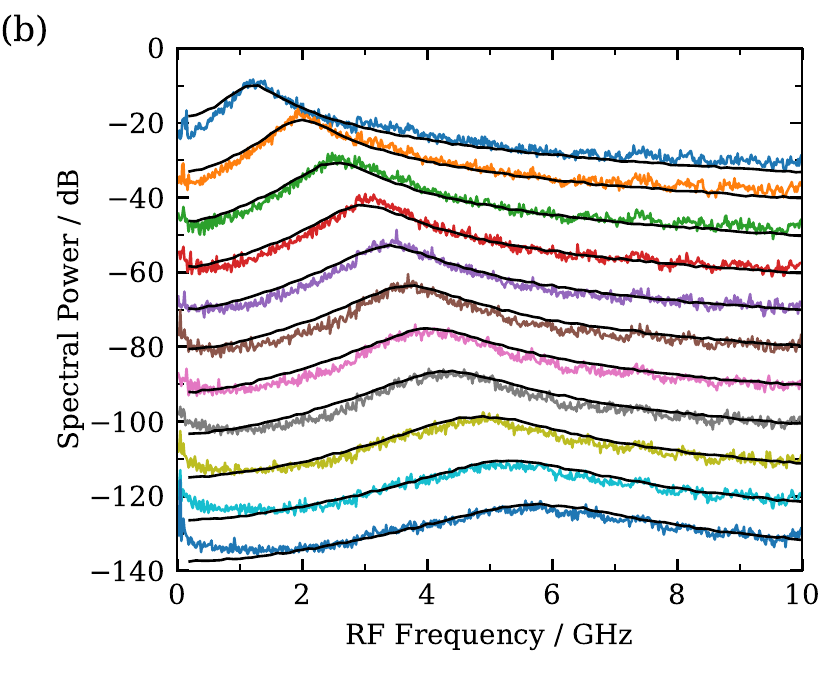}
 \caption{\textbf{(a)} Calculated RF spectrum of a slave laser under optical injection from a $10~\GHz$ comb, as the pump current in the slave is swept. The dashed white line indicates the frequency of the free-running ROs. The  injection strength of $K = 0.06$ was fixed. \textbf{(b)} Measured free-running ROs of the slave laser from $55~\mA$ to $120~\mA$, in steps of $5~\mA$ up until $90~\mA$ and steps of $10~\mA$ thereafter, offset vertically from top to bottom by $-10~\dB$ for clarity. Solid black lines indicate fits from which we determined parameter values for our model.}
 \label{Fig_jsweep_with_RO_fits}
\end{figure}

With this first example, it is already clear that the undamping of the ROs and their subsequent frequency locking depends on the relative detuning between the injected comb and the slave. Not surprisingly, the relative optical power injected is also important. However the relationship between the RO frequency and the optical comb spacing is not immediately clear.
The ROs can in fact lock to any rational fraction $p/q$ of the optical comb's frequency spacing. To demonstrate 
this, in Fig.~\ref{Fig_jsweep_with_RO_fits}(a) we simulate the RF spectrum of a slave laser optically injected by a three line $10~\GHz$ comb, as the current in the slave laser is swept numerically. In the calculations, the detuning between the slave and comb was fixed at zero, with constant relative injection strength $K = 0.06$. 
The analytically calculated RO frequency of the slave laser (dashed white line) shows the free-running behaviour of the slave. While in the free-running laser the ROs are damped and stochastically excited by spontaneous emission and carrier noise (cf. Fig.~\ref{Fig_jsweep_with_RO_fits}(b)), under comb injection the ROs are clearly undamped, exhibiting pronounced narrow peaks in the RF spectrum. This RF peak closely follows the analytically calculated RO frequency dependence. However, several regions of frequency locking can be observed, where the RO frequency remains at a constant value under change of the pump current. Several such locking regions can be observed, with a small region at $2.5~\GHz$ above threshold, at $3.33~\GHz$ near $J=1.4J_\th$, and at $5~\GHz$ around $J=2J_\th$. These locking regions correspond to a harmonic locking to the comb frequency spacing, at $\tfrac14$, $\tfrac13$, and $\tfrac12$ of the comb spacing. The RO locking bandwidth is wider for a smaller denominator $q$ of the rational fraction at which it locks. We have experimentally demonstrated RO locking for ratios of $\tfrac14$ of the comb spacing and above, however smaller fractions have not yet been realised due to the narrower locking ranges involved.

We can determine from our model that the RO frequency locking is enabled by an intensity modulation caused by the beating of the undamped ROs with the neighbouring unlocked comb lines. The resulting periodic forcing from this beating is analogous to the periodic forcing as seen in the circle map, where locking tongues can be observed at rational ratios of the driving frequency \cite{Arnold1965, Lingnau2019}. As a result, this effect is strongest when the slave is locked to the centre line of the optical comb, due to the symmetry of the unlocked comb lines.

\begin{figure*}
\centering
  \includegraphics[width=1.99\columnwidth]{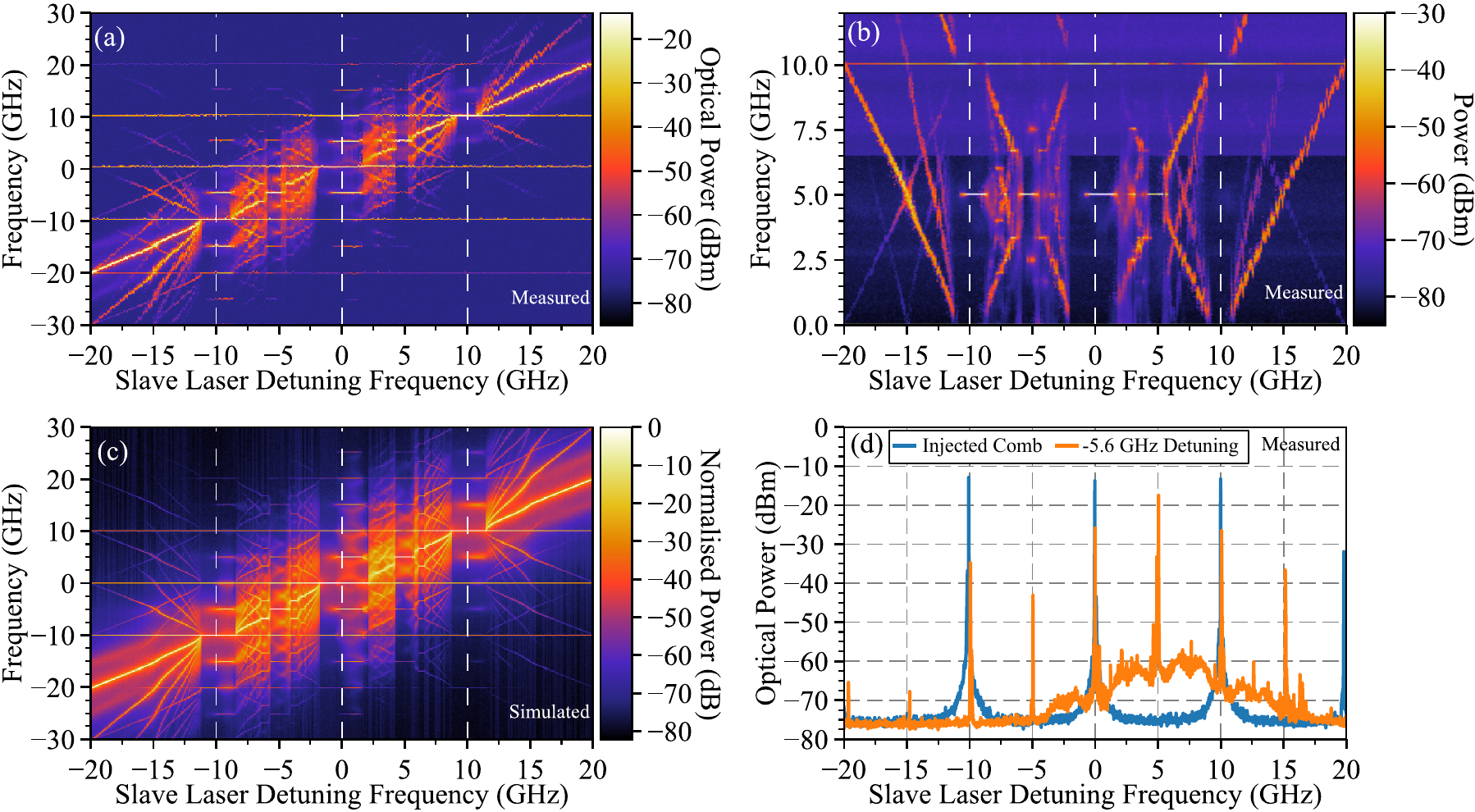}
 \caption{ 
Experimental and theoretical results from a slave laser at $1.75$ times threshold current with free-running power -4.5dBm, injected by 3 line comb of total power -7dBm, with spacing $\nu\cdot\kappa=10~\GHz$.  The vertical dashed white lines show where the free-running laser frequency is resonant to one of the optical comb lines. \textbf{(a)} Intensity plot of measured optical spectra as the slave laser was tuned across the injected optical comb. The frequency axis has been scaled such that $0~\GHz$ coincides with the centre comb line.  \textbf{(b)} Electrical power spectrum for the corresponding frequency sweep in (a).
\textbf{(c)} Intensity plot of simulated optical spectra with $K=0.06$, for the same parameters as in (a).  \textbf{(d)} Comparison between the optical spectra of the injected comb, and the output comb as the slave is frequency locked between the optical comb lines (at a detuning of $-5.6~\GHz$).}
 \label{fig:10GHzDetSweep}
\end{figure*}

Figure~\ref{fig:10GHzDetSweep} presents results from a detuning sweep, as a single mode slave laser's frequency is tuned from -20 GHz to +20 GHz
relative to the centre line of a 3 line $10~\GHz$ comb. In Fig.   \ref{fig:10GHzDetSweep}(a), an intensity plot shows the output optical spectrum of the slave laser. Vertical dashed lines indicate where the frequency of the slave matched that of one of the comb lines. On the left of each of these three dashed lines, the slave laser undergoes standard Adler-type frequency locking, its frequency entrained by the closest optical comb line. As the detuning positively increases however, the ROs in the slave laser become undamped and then lock to a $5~\GHz$ frequency, generating additional optical frequencies in the slave laser's spectrum (as shown in the example in Fig.~\ref{fig:RO_Excitation_Plots}). Further detail on the frequencies present in the detuning sweep is visible in the ESA data in Fig.~\ref{fig:10GHzDetSweep}(b). As the slave laser undergoes Adler-type locking, the electrical spectrum shows only the strong $10~\GHz$ frequency and its harmonics (not shown). This $10~\GHz$ frequency extends across all detuning values, due to the beating of the injected comb lines. Sub-harmonics at $5~\GHz$ are generated when the ROs become excited, and undergo a second 
frequency locking. The sub-harmonics lock to exactly half the comb spacing, then  
subsequently unlock and split as the detuning increases, prior to the slave laser becoming unlocked entirely from the comb line.

A second type of non-adler locking is observed as the slave laser is tuned between comb lines, with specific resonant frequencies appearing in Fig.~\ref{fig:10GHzDetSweep}(b). As a result, beat notes 
at $2.5~\GHz$, $3.3~\GHz$, $5~\GHz$ and their higher harmonics are visible on the ESA spectrum. In particular, the locking bandwidth of the $5~\GHz$ resonances (or, the $1:2$ resonances), located at $\pm 5~\GHz$ detuning, are the largest (although this is not necessarily always the case). As detailed in Ref.  \cite{Lingnau2019, Tistomo2011}, the locking structure as the slave tunes between comb lines is very well approximated by a devil's staircase, with all frequencies at rational fractions of the injected comb's frequency spacing present. Again, this locking occurs due to the periodic forcing caused by the neighbouring comb lines. As the slave's frequency becomes resonant with a rational fraction of this periodic forcing, it becomes frequency locked. These resonances are more commonly known as Arnol'd tongues.

Figure~\ref{fig:10GHzDetSweep}(d) shows the optical spectrum at a detuning of $-5.6~\GHz$, in comparison with the injected comb. Although the slave laser can in principle lock to any resonance order, the variation in the power level of the tones in the comb vary more than in the RO locked case. Nevertheless, this non-Adler type locking is stable and could therefore be of technological interest. A further example of a frequency detuning sweep is presented in Appendix A, which discusses the asymmetry of excitations with respect to the central comb line.

Using our model, we reproduce the same detuning sweep, and the simulated results from the OSA spectrum are shown in Fig.~\ref{fig:10GHzDetSweep}(c). The calculated slave laser frequency sweep accurately reproduces the main characteristics in both the OSA and ESA (not shown) experimental data. Each of the three Adler locking events in Fig.~\ref{fig:10GHzDetSweep}(c) initially lead to a fundamental locking to the comb, and then subharmonics can be seen for slightly higher detuning values. 
The structure found as the slave laser is tuned between the comb lines is also well replicated, with the simulated slave locking to the same resonances between the comb lines. The agreement seen between theory and experiment reinforces our understanding of how the unlocked comb lines interact with the slave laser; in the following section, some of the model parameters are varied to investigate their role in these types of locking.

\section{Relative Locking Stability and the Role of Linewidth Enhancement Factor}
\label{sec:Stability_and_Alpha}
With strong agreement between theory and experiment, our model can be used to further understand the stability of the frequency and RO  locking, and how the $\alpha$ parameter affects the size of the locking tongues.

\begin{figure*}
\includegraphics[width=0.99\columnwidth]{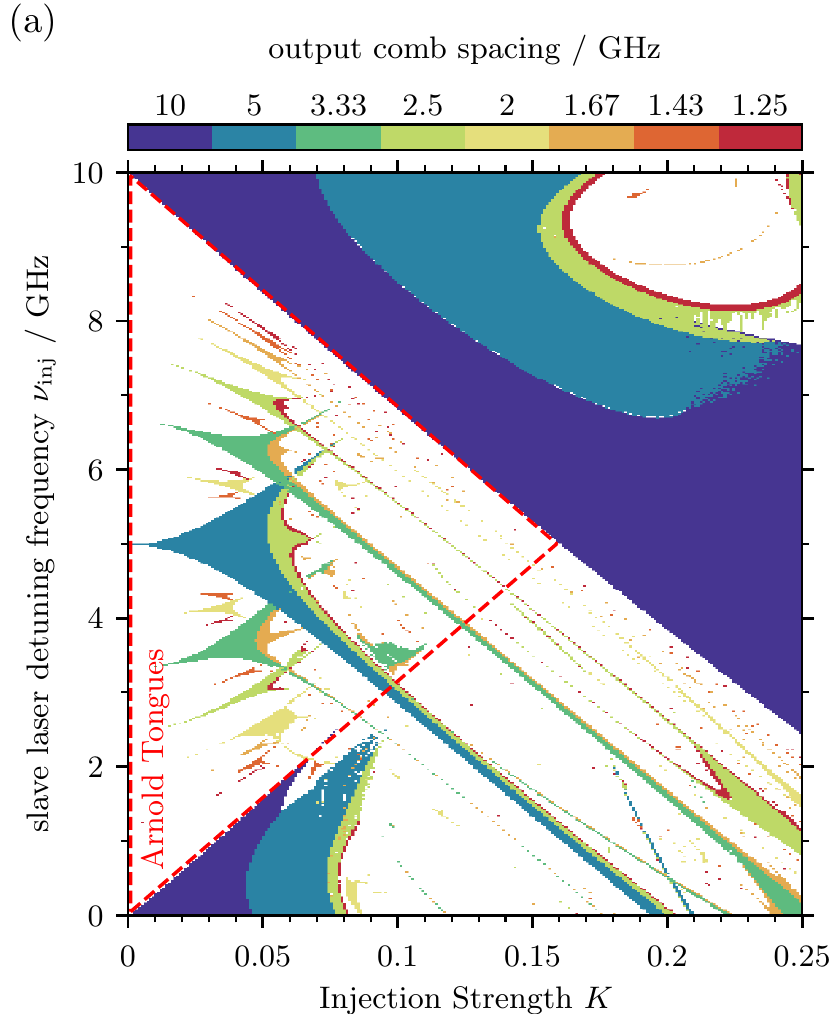}
\includegraphics[width=0.99\columnwidth]{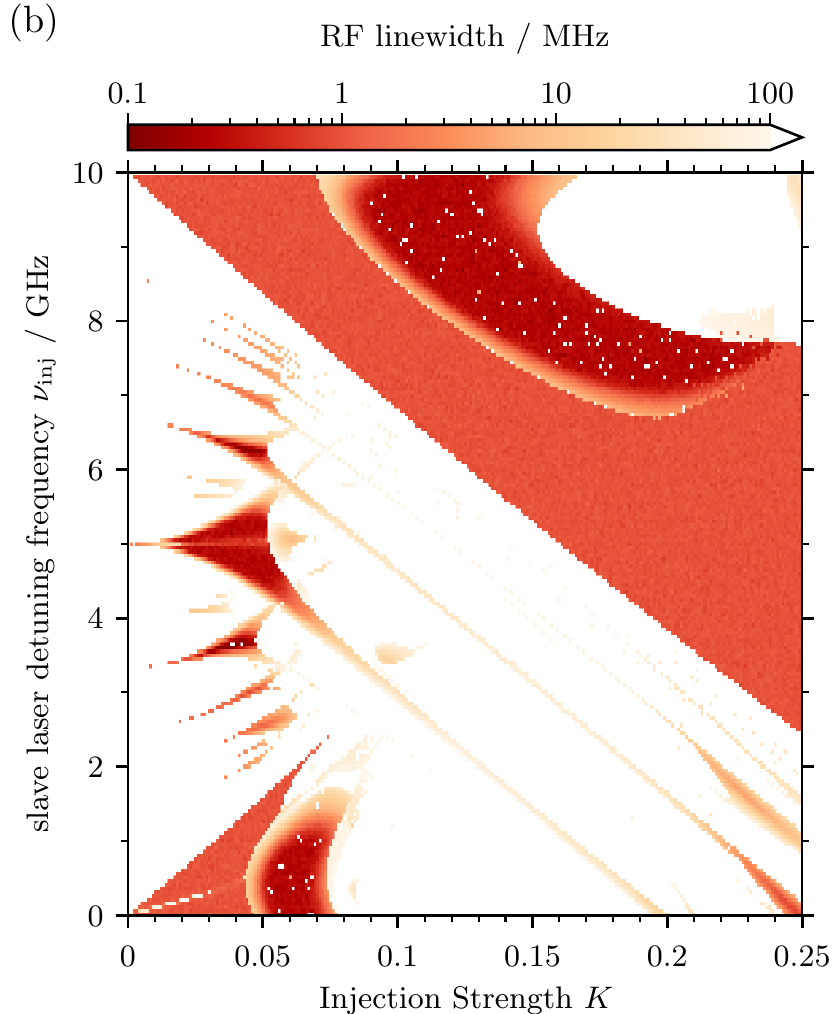}
 \caption{Two dimensional maps of the parameter space spanned by detuning and injected power, calculated with $J = 1.82J_\th$  and $\Delta \cdot \kappa = 10~\GHz$. \textbf{(a)} Map of the output comb frequency spacing. White regions indicate unlocked states. The triangular region enclosed within the dashed lines shows the parameter region used to calculate the relative size of the harmonic locking areas in Fig. \ref{fig:Region_Counts}. \textbf{(b)} RF linewidth of the dominant harmonic's beat note, over the same parameter space as in (a).}
\label{fig:Harmonics_and_Stability}
\end{figure*}

To study the relative range of parameter values that leads to the observed types of locking, two dimensional maps of the parameter space spanned by detuning $\nu_\inj$ and optical injection strength $K$ were calculated and are shown in Fig.~ \ref{fig:Harmonics_and_Stability} for a three line optical comb with $10~\GHz$ spacing. In Fig.~ \ref{fig:Harmonics_and_Stability} (a), 
the output comb spacing from the slave laser is monitored, for positive detunings only.  
Large Adler-type locking cones are centred around $0~\GHz$ and $10~\GHz$ detuning, where the slave laser is locked to one of the main comb lines.
As the injection strength is increased within these tongues, we see the output comb spacing half in a period-doubling bifurcation, leading to a significantly large area of harmonic locking within the main locking cones. 
As demonstrated above in Section  \ref{Sec:Freq_Locking}, within these regions the slave laser's ROs become undamped and lock to one half of the optical comb's spacing (the ROs lock to the $5~\GHz$ subharmonic in this case, as the pump current used was $J=400\approx1.8 J_\th$, cf. Fig.~\ref{Fig_jsweep_with_RO_fits} (a)).
The map also shows asymmetry between locking to the centre or higher frequency lines in the comb, as the RO locking occurs at lower injection strengths when the slave is locked to the centre comb line.
Interestingly, we see further period doubling at even higher injection strengths in both tongues, before the slave unlocks from the injected comb. 
In order to characterise the stability of each of the locked states, the RF linewidth of the dominant RF peak in each state was calculated over the same parameter space and is presented in Fig.~ \ref{fig:Harmonics_and_Stability}(b). 
At low injection strengths, the Adler locking tongues around $0~\GHz$ and $10~\GHz$ detuning have RF linewidths close to the simulated comb linewidth of $1~\MHz$, suggesting these are stable locked states as expected.  Within the harmonic RO locking regions of these tongues, the RF linewidth decreases further to below $1~\MHz$, thus improving the comb stability even beyond that of the driving optical comb. 
The boundary of the locked RO region is unstable however, as the noise stochastically kicks the injected laser out of the harmonic locking state, leading to a broad RF peak. Higher harmonic locking beyond the secondary period doubling bifurcations at even higher injection strengths can be seen to be highly unstable, with linewidths in excess of $100~\MHz$, suggesting a clean comb output can be obtained only for harmonic locking ratios that are not too high.


Smaller Arnol'd-type locking tongues can be seen in between the main Adler locking tongues in Fig.~\ref{fig:Harmonics_and_Stability}(a). First reported in studies of the circle map \cite{Arnold1965, Brøns1997, Ecke1989}, these harmonic locking tongues appear due to the periodic forcing of the unlocked comb lines on the slave laser \cite{Lingnau2019}. For low injection strengths, Arnol'd tongues can be seen at detunings corresponding to rational fractions $p/q$ of the optical comb spacing. The $1:2$ resonance tongue, located at $5~\GHz$ for lower injection powers, is the largest of these Arnol'd tongues, locking the frequency of the slave laser mid-way between the comb lines, causing the output comb spacing of the slave to be $5~\GHz$. Within this locking tongue, we see some period doubling occur as the injection power is increased. The RF linewidth map in Fig.~\ref{fig:Harmonics_and_Stability}(b) suggests that the centre of the $1:2$ resonance Arnol'd tongue is stable, with unstable boundaries. As with the Adler tongues, the period doubled states seen as the injection strength is increased within the $1:2$ resonance tongue are very unstable. Figure~ \ref{fig:Harmonics_and_Stability}(a) clearly shows the sizes of the $1:3$, $1:4$ and $1:5$ resonances shrink as the denominator in the rational fraction grows. The corresponding RF linewidth calculation also suggests that these resonances become less stable as the denominator grows.


\subsection{Relationship Between the Relaxation Oscillations and the Size of Arnol'd Tongues}
\label{ROs_and_Arnold}

\begin{figure*}
\includegraphics[width=0.99\columnwidth]{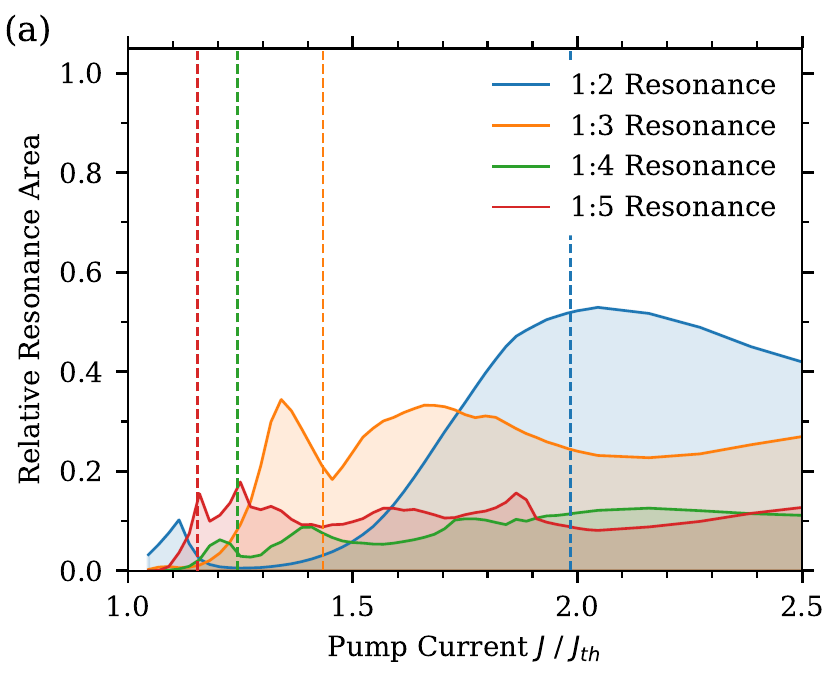}
\includegraphics[width=0.99\columnwidth]{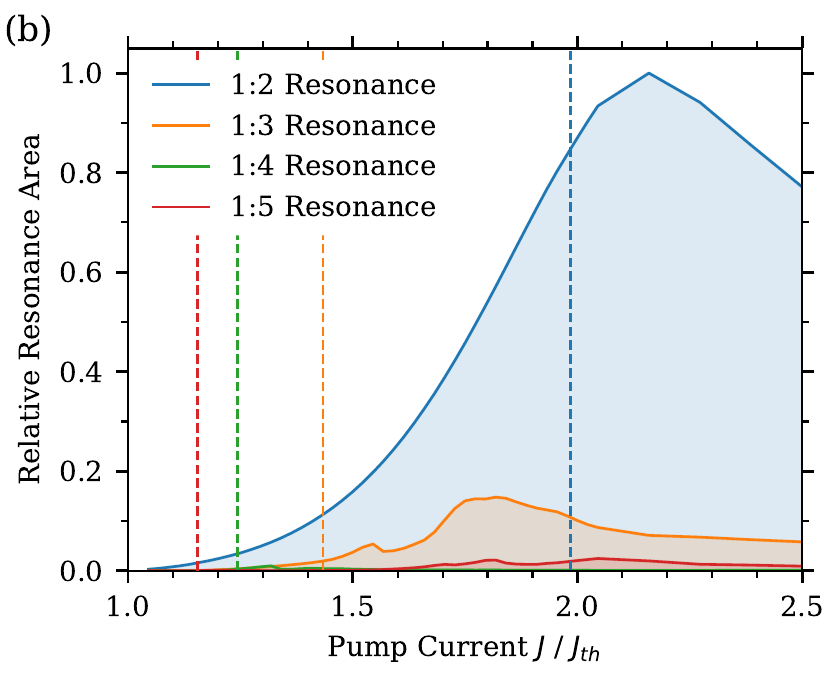}
 \caption{ Plots showing the relative size of the Arnol'd resonances as a function of slave laser pump current, for \textbf{(a)} $\alpha=3$, and \textbf{(b)} $\alpha=0$. For each current, a two dimensional map over injection strength and detuning was calculated as in Fig.~ \ref{fig:Harmonics_and_Stability}, and only the area contained in each resonant Arnol'd tongue was counted. The vertical dotted lines indicate the pump currents at which the free-running RO frequency matches that of a resonance.}
\label{fig:Region_Counts}
\end{figure*}

As presented above in Fig.~ \ref{Fig_jsweep_with_RO_fits}, the free-running frequency of the ROs has to be close to a rational fraction of the comb's frequency spacing, in order for the ROs to lock. The size of the locked RO regions within the Adler-type locking tongues in Fig.~\ref{fig:Harmonics_and_Stability} can thus be expected to depend critically on  how close the free-running ROs are to a subharmonic of the comb spacing. 
In the following, we investigate the dependence of the Arnol'd regions on the RO frequency of the free-running laser. Figure~ \ref{fig:Region_Counts} (a) shows a measure of the area of each Arnol'd resonance tongue in the parameter space within the red dashed triangle in Fig.~\ref{fig:Harmonics_and_Stability} (a) as a function of the slave pump current, for $\alpha = 3$. 
For each pump current, a two dimensional map was computed, and the size of each Arnol'd tongue within the area enclosed in the dotted red triangle in Fig.~ \ref{fig:Harmonics_and_Stability} (a) was counted. The vertical dotted lines in Fig.~ \ref{fig:Region_Counts}(a) indicate the pump currents at which the free-running RO frequency matches that of a resonance. The results show a strong correlation between the size of the frequency locked resonances and the RO frequency. 
We find that the $1:2$ resonance only becomes the largest resonance as the RO frequency approaches half the injected comb spacing. Likewise, as the RO frequency approaches one-third and one-fifth of the comb spacing, the $1:3$ and $1:5$ resonances are the largest in the parameter space. The same is not true for the $1:4$ resonance.
Interestingly, in each case, the maximum of the resonance area doesn't correspond exactly with the pump current where the ROs frequency exactly matches the resonance. 
This could be in part due to variance between the free-running ROs and the ROs in the locked slave laser\cite{Petitbon1988}. The ROs in an injection locked slave laser also vary with optical injection strength and detuning \cite{Mogensen1985}, both of which are parameters varied in the two dimensional resonance areas measured. Nevertheless, the theoretical predictions suggest that for a specific harmonic resonance an optimal pump current exists at which that resonance can be produced reliably.

Figure~ \ref{fig:Region_Counts} (b) repeats the calculation of Fig.~ \ref{fig:Region_Counts} (a) with $\alpha = 0$. It is immediately clear that the magnitude of the amplitude-phase coupling in the slave laser's electric field has a strong impact on the how large the resonance areas are. Interestingly, the $1:2$ resonance grows in size with decreasing $\alpha$, with the $1:2$ resonance remaining dominant for all pump current values in this case. The higher order resonances are severely suppressed in the $\alpha = 0$ case. It is also noteworthy that the correlation between the free-running RO frequency and the peaks of the resonance areas is much weaker in this case, showing that the nonlinearity in the laser response induced by $\alpha$ is a main driving mechanism for the excitation of the higher order resonances.

\begin{figure*}
\includegraphics[width=1.99\columnwidth]{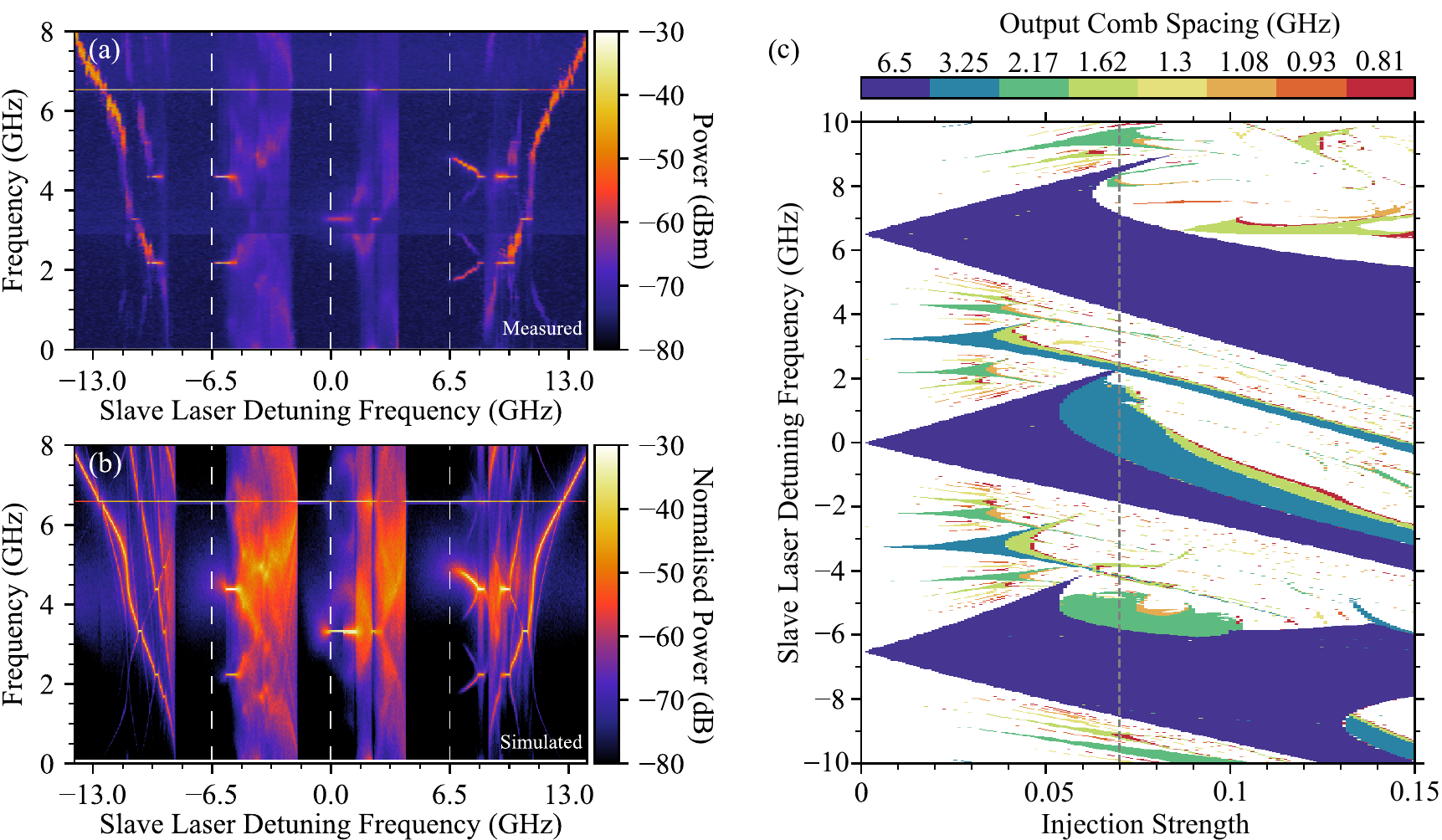}
 \caption{ \textbf{(a)} Intensity plot showing the  electrical spectra measured as the frequency of a single mode slave laser was swept the injected $6.5~\GHz$ optical comb. The optical comb had a power of $-5.0~$dBm, while the slave laser was fixed at $1.75$ times its threshold current, with $-4.4~$dBm coupled to the lensed fibre. The vertical dashed white lines show the parameter values where the free-running laser frequency is resonant to one of the optical comb lines. \textbf{(b)} Corresponding simulated experiment for the parameters in (a). \textbf{(c)} Two dimensional map  of the parameter space spanned by detuning and injected power, showing the output comb frequency spacing, calculated with $J = 1.75J_\th$  and $\Delta \cdot \kappa = 6.5~\GHz$. The vertical dotted line indicates the injection strength used in (b). White regions indicate unlocked states.}
\label{fig:6.5GHz_Example}
\end{figure*}

\section{Conclusion}
In conclusion, we have presented an experimental and numerical study on two types of frequency locking which occur with optical comb injection. Namely, the locking of the undamped relaxation oscillations, and the Arnol'd type higher order resonance tongues when the slave laser's frequency lies between the comb lines. In effect, both types of locking can be used to decrease the optical comb spacing of the original injected optical comb. We have shown that relaxation oscillations of the injected laser can lock to harmonics of the injected optical comb, generating extra tones in the optical comb around the slave laser's lasing frequency. The relaxation oscillations can in principle lock to any rational fraction of the input comb's frequency spacing. However we showed in simulations that the extent of the locking range in general decreases with the order of the harmonic resonance. 
Furthermore, through analysing the simulated RF linewidth, we showed that both Arnol'd frequency locking and the relaxation oscillation locking can be as stable as the standard Adler-type locking, and in some cases reduce the RF linewidth of the produced output comb compared to the injected comb. The effect of the free-running relaxation oscillation frequency on the extent of the Arnol'd resonances in parameter space was also discussed.
We have shown that the locking range for producing a given harmonic comb can be optimised by tuning the slave laser relaxation oscillation frequency to an optimal value via the pump current. We have furthermore shown the nonlinearity induced by the $\alpha$ parameter to be a driving mechanism for the generation of harmonic locked states in the injected slave laser.

The presented locking scenarios can be used to generate optical combs with variable frequency spacing, as well as regenerate noisy frequency combs via the nonlinear interaction within the laser device, producing a comb with narrower RF linewidth.

\section*{Funding}
Science Foundation Ireland (SFI 13/IA/1960); Deutsche Forschungsgemeinschaft (DFG, German Research Foundation) (404943123).


\appendix

\section{Asymmetry of Relaxation Oscillation Locking around the Centre Comb Line}
\label{App_Sweep_Examples}

In this appendix, we present further results detailing the asymmetry of the RO locking around the optical comb centre. The results which motivate this discussion are presented in Fig.~\ref{fig:6.5GHz_Example}.  Similar to the detuning sweeps presented above, Fig.~\ref{fig:6.5GHz_Example}(a) and (b) show measured and simulated ESA spectra as the frequency of a slave laser was tuned $\pm14~\GHz$ (from negative to positive) relative to the centre of a $6.5~\GHz$ comb. In this case, an optical comb spacing of $6.5~\GHz$ was chosen to ensure that the free-running ROs of the slave laser approximately matched $\tfrac23$ of the comb spacing. 

While locked to the lower frequency comb line, the slave laser ROs become undamped and lock to $\tfrac13$ of the comb spacing. The bandwidth over which the ROs remain locked is narrow, relative to the $10~\GHz$ case shown previously. While the slave is locked to the centre line however, we note that the ROs lock instead to $\tfrac12$ of the comb spacing. This highlights the asymmetry introduced by using a 3 line optical comb -- the modulation effect is stronger at zero detuning as the neighbouring comb lines have higher intensity, and as a result the modulation is strong enough to force the ROs to lock to a frequency of $3.25~\GHz$. For the higher frequency comb line, the ROs become undamped and instead of locking immediately, tune in frequency from $4.8~\GHz$ to $4.33~\GHz$. The difference between the behaviour of the ROs while locked to the lower frequency and higher comb lines is due to the asymmetry introduced by the amplitude-phase coupling. 

Figure \ref{fig:6.5GHz_Example}(c) shows a two dimensional map which highlights the output comb spacing, for the parameter space spanned by detuning $\nu_\inj$ and optical injection strength $K$.
The vertical dotted line indicates the injection power used in (b). In this case, we notice that the Arnol'd tongues are significantly smaller in size than previously seen in the $10~\GHz$ case, due to the narrower spacing of the optical combs. This is reflected in both the experimental and simulated detuning sweeps, as neither have strong locking events as the slave tunes between the comb lines. Likewise, the two dimensional parameter map shows different RO locking behaviour within the three Adler locking tongues. The lower frequency Adler tongue has a region where the ROs lock to $\tfrac13$ of the optical comb spacing. In the centre Adler tongue, the ROs lock instead to $\tfrac12$ of the input comb spacing, and as seen in experiment, the ROs don't lock at all in the higher detuning tongue. We expect however that as the number of comb lines increases, the behaviour within the Adler locking tongues will only differ significantly for the outer comb lines.

\end{document}